\documentclass{iaus}
\usepackage{graphicx}

\title[Cosmic Recycling: PNe Abundances] 
{Reduce, Reuse, Recycle: Planetary Nebulae \\as Green Galactic Citizens}

\author[Karen B. Kwitter \& Richard B. C. Henry]   
{Karen B. Kwitter$^1$
 \and Richard B. C. Henry$^2$}

\affiliation{$^1$Astronomy Department, Williams College, \\ 33 Lab Campus Drive, 
Williamstown, MA 01267 USA \\ email: {\tt kkwitter@williams.edu} \\[\affilskip]
$^2$H. L. Dodge Department of Physics and Astronomy, University of Oklahoma, \\ Nielsen Hall, 
Norman, OK 73019 USA \\email: {\tt henry@nhn.ou.edu}}

\pubyear{2011}
\volume{283}  
\pagerange{}
\jname{Planetary Nebulae: An Eye to the Future}
\editors{A. Manchado \& L. Stanghellini, eds.}
\begin{document}

\maketitle

\begin{abstract}
We review gas-phase abundances in PNe and describe their dual utility as
archives of original progenitor metallicity via the $\alpha$ elements, as well
as sources of processed material from nucleosynthesis during the star's
evolution, i.e., C, N, and s-process elements. We describe the analysis of PN spectra to derive abundances 
and discuss the discrepancies that arise from different choices at each step. 
Abundance results for the Milky Way and Magellanic Clouds from various groups of investigators are presented;  the observational results are compared with theoretical predictions of AGB stellar yields. Finally, we suggest areas where more work is needed to improve our abilities to determine
abundances in PNe.
\end{abstract}

\firstsection 
\section{Introduction}
Chemical abundances in planetary nebulae (PNe) offer a window into the nucleosynthesis and interior processes occurring in low- to intermediate-mass stars (LIMS). Abundances in individual PNe illuminate this history star by star; combined results for a galaxy's ensemble of PNe provide a broader view of chemical evolution. In particular, the products of nuclear burning and convective dredge-up episodes depend on properties of the specific PN progenitor, while the $\alpha$ elements, largely unaffected by LIMS nucleosynthesis, archive the progenitor's original abundances and can reveal galaxy-wide characteristics. In this review, we describe the observational data and calculation methods used to derive PN abundances, discuss uncertainties in the process, and present recent abundance results, comparing them with theoretical predictions from AGB models.

\section{Observations}
Long-slit, moderate-dispersion optical spectroscopy has been the mainstay of PN abundance determinations for decades. The resulting 1-D spectra -- sometimes divided into sub-spectra along the slit to yield a modicum of spatial resolution -- have provided abundance data for hundreds of PNe in the Milky Way (MW) and Magellanic Clouds (e.g., see references to Table 1). The last decade has seen the rise of integral field spectroscopy, in which each pixel yields a spectrum, rather than a single intensity. Such multiplexed spectroscopy has allowed 2-D mapping of nebular properties (e.g., \cite{T08}) and provided data that can be used to construct and test theoretical PN models.

Beyond the optical region, {\it Spitzer} has enabled us to detect infrared lines from important ions (e.g., [S~IV] , [Ar~II], and [O~IV]) that are otherwise invisible; {\it Hubble} has done the same in the ultraviolet (e.g., N~III], [N~V], and C~III]). Expanding the wavelength regime of observed ions reduces dependence on {\it ionization correction factors}, (ICFs), yielding more accurate abundances, discussed further in \S 3, below. 

\section{Data Analysis}
In the absence of a customized photoionization model, abundances in a PN are typically calculated empirically, using a multi-level atom program such as {\it nebular} (\cite{SD95}), or {\it ELSA} (\cite{J06}). While space does not permit a full examination of all relevant issues in data analysis, we  briefly discuss the three main steps, each of which presents opportunities for different choices that will affect the results.
\subsection{The Empirical Method}
After removal of the instrumental signature from the observed fluxes (itself a possible source of systematic error), a wavelength-dependent correction for interstellar extinction is applied. There are many versions of this function; among those currently in use are \cite{CCM}, \cite{F99}, and \cite{SM}. De-reddened line flux ratios will differ systematically depending on which function is chosen.

Transforming de-reddened flux ratios into ionic abundances requires knowledge of atomic data, including recombination rates for permitted lines and A-values and collision strengths for collisionally-excited lines. Data cited by \cite{M83} have been widely used over the past several decades; Liu (this volume) describes some more recent calculations. Use of different atomic data can produce large discrepancies in the derived abundances (see \S4.2). An important issue that needs to be addressed as we go forward is that when new atomic data are published, it is not easy for the end-user to evaluate their quality and validity; also, ``the newest data are not always the best" (see Luridiana \& Garc\'{i}a-Rojas, this volume). 

To convert ionic abundances into total elemental abundances, one must account for unseen ionization stages. Without a tailored model, one must use {\it ionization correction factors} -- the ICF is the ratio of the total elemental abundance to the sum of the detected ionic abundances. Most ICFs exploit coincidences in ionization potentials with ions of oxygen and helium, the most thoroughly-observed elements; in some cases (e.g., S), ICFs come from modeling results (see \cite{KH01}). \cite{KB94} published a set of ICFs that remains in wide use. Despite the {\it ad hoc} nature of ICFs, they work well in most circumstances. However, there are instances where they do not, as for Ne under low-excitation conditions, leading to abundances that are too low (see \cite{P92}; Henry et al. 2011 {\it priv. comm.}).

The bottom line is that at each step outlined above, the particular choice made will affect the final abundances, and can potentially lead to significant disagreements compared with results by others who have made different choices, {\bf even if they started with identical measured fluxes}. It is therefore desirable for those who work in PN abundances to make their choices public -- and perhaps even to discuss them collaboratively -- in order to disperse the chaff unrelated to true abundance behavior in PNe.

\subsection{The Abundance Discrepancy - still unresolved}
When it became possible to measure weak recombination lines from light elements, it was discovered that they implied abundances in excess of those derived from collisionally-excited lines (e.g., \cite{T03}), sometimes by large factors. Proposed explanations invoke chemical inhomogeneities in the form of cold, H-deficient inclusions (e.g., \cite{L06}) or temperature and density variations in a chemically homogeneous nebula (e.g., \cite{TPP03}). Challenges to the resolution of this issue have included the lack of atomic data relevant to the low temperatures assumed for these inclusions (though this is being addressed --Ê see Liu, this volume), and the absence of definitive observational data. Since the proposed inclusions would radiate strongly in infrared fine-structure lines, the spatially-resolved SOFIA observations of NGC~7009 described by Rubin (this volume) may prove useful.

\section{Light Element Abundances in the Milky Way, LMC, and SMC}
\subsection{Milky Way Disk, Bulge and Halo}
In Table~1 we present a compilation of recent abundance results for He, C, N, O, Ne, S, and Ar in the MW disk, bulge and halo, and for the Magellanic Clouds. All the MW disk abundances agree within the uncertainties. The $\alpha$ element abundances are close to solar, while He, N and C are enhanced. S appears underabundant; this {\it sulfur anomaly}, has long been a problem to understand (Henry et al. 2004, Henry et al. (this volume)).

Bulge PNe exhibit He, O, Ne, and Ar abundances similar to the disk, while C and N appear slightly less abundant, and the S anomaly remains. The halo appears deficient compared with the disk in all elements measured, except for C. It should be noted that the halo PN sample is small and diverse; the median (or average) abundances are therefore not as robust an indicator of overall behavior as they are for other samples.

\begin{table}[h]
\begin{center}
\caption{Abundance Comparisons (median 12+log(X/H))}
\label{tab1}
{\scriptsize
\begin{tabular}{lcccccccc}\hline

{\bf PN Sample} & {\bf He} &{\bf C}&{\bf N}&{\bf O}&{\bf Ne}&{\bf S}& {\bf Ar}& {\bf Ref.}\\ \hline
MW disk	&	11.09&	8.87& 	8.33&	8.61&	8.02&	6.69&	6.40&	1 \\
		&	11.04 &	& 	8.13& 	8.61& 8.00	& 	& &	2\\
		&	11.03&	8.64	&	8.18&	8.62&	8.13&	6.87& 6.43&	3\\
MW Bulge&	11.05	&8.44&8.10&	8.71	&	8.03&	6.99&	6.26&	4\\	
		&	11.11	&&8.11&	8.57	&	7.93&	6.79&	6.34&	5\\	
		&	11.04&	&	8.01&	8.48&	7.79&	6.59&	6.37&	6\\	
MW Halo	&	10.97&	9.00&	7.54&	7.83&	6.82&	5.39&4.78&	7\\
		&	10.99&	&	7.76&	8.46&	7.81&	&	&		2\\
LMC		&	11.00&8.17	&	7.75&	8.24&	7.62&	7.08&	5.90&	5; 8 for C\\
		&	11.01&	8.56&	7.79&	8.31&	7.56&	7.27&	6.00&	9\\
		&	&	&	&	&	7.75&	6.47&	&	10\\
SMC		&	11.11&	8.25	&	7.17	&	8.04&	7.40&	6.11&	5.57&	11; 12 for C\\
		&	10.95&	&	7.26&	8.10&	7.35&	6.72&	5.63&	13\\
		&	&	&	&	&	7.45&	6.04&	&	10\\			
Orion	&	10.99&	8.52&	7.73&	8.73&	8.05&	7.22&	6.62&		14\\
Sun		&	10.93&	8.43&	7.83&	8.69&	7.93&	7.12&	6.40&		15\\
\hline

\end{tabular}
}
\end{center}
\vspace{1mm}
 \scriptsize{
 {\it References:}
  1: \cite{H10}; 2: \cite{SH10}; 3: \cite{PBS10}; 4: \cite{WL07}; 5: \cite{Ch09}; 6: \cite{Ca10}; 7: \cite{HKB04}; 8: \cite{S05}; 9: \cite{LD06}; 10: \cite{BS08}; 11: \cite{Shaw10}; 12: \cite{S09}; 13: \cite{IMC07}; 14: \cite{esteban04}; 15: \cite{asplund09}
 }
\end{table}

\subsection{The Magellanic Clouds}
As has long been known, the Magellanic Clouds are less chemically evolved than the MW. Both Clouds are deficient in He compared with the MW disk, the LMC by $\sim$0.05 dex, and the SMC by $\sim$0.1 dex. The SMC He abundance from \cite{Shaw10} is 0.15 dex higher than that of Idiart et al. (2007), arising from the use of a new He$^+$ recombination rate (\cite{P07}). This is a perfect example of why it is imperative to understand how results are obtained by different groups before forming conclusions about any discrepancies.  The median N abundances in the LMC and SMC are below the MW disk by $\sim$0.5 dex and $\sim$1 dex, respectively. The LMC and SMC median C abundances are similar to each other, presumably due to the lack of hot-bottom burning in the SMC PNe progenitor stars. The abundances of the $\alpha$ elements are lower in the LMC than in the MW disk by $\sim$0.4 dex, and in the SMC by $\sim$0.6 dex, except for Ar, which is lower in the SMC by $\sim$0.8 dex (though shortcomings in the ICF for argon may be a contributing factor).

\section{Abundances of s-process Elements and Iron in Milky Way PNe}
The s-process elements are produced in the interpulse phase of the TP-AGB in the region between the hydrogen- and helium-burning shells; third dredge-up (TDU) then carries them, along with $^{12}$C, to the surface. Since the first report of s-process elements in PNe by \cite{PB94}, much progress has been made, especially in the last several years.  \cite{Sharpee} detected lines of Br, Kr, Xe, Rb, Ba, and Pb and possible lines of Te and I, and found that  abundances of Kr and Xe were enhanced. \cite{SD08}(SD08) detected Kr and/or Se in 81 of 120 PNe; in non-Type~I PNe they found average enhancements of $\sim$0.3 dex for Se, and $\sim$1 dex for Kr. Type I PNe exhibited little s-process enrichment, which SD08 attribute to smaller intershell masses and efficient dilution of s-process material in the massive AGB envelopes. They found a positive correlation between s-process enrichments and C/O, as expected in TDU. Future progress in learning more from s-process abundances in PNe relies heavily on knowing transition rates and collision strengths (See Sterling, this volume).

\cite{D09} observed [Fe~III] lines in 33 low-ionization PNe. They found a median 12+log(Fe/H) = 5.85, and a range in depletion, [Fe/H], from -1.01 to  -3.2, implying depletion $>$ 90\%, and M$_{dust}$/M$_{gas}$ $\geq$ 1.3 x 10$^{-3}$. Further, they found that the depletion correlates somewhat with C/O ratio (see Delgado Inglada, this volume).

\section{Abundance Correlations and Model Tracks}
\subsection{Helium, Carbon, Nitrogen \& Oxygen}
Fig.~1 shows N vs. He for a variety of PN samples. The MW disk PNe are taken from \cite{HKB04}, reanalyzed with consistent atomic data; these PNe all have good measurements of T$_{[OIII]}$, T$_{[NII]}$, and N$_{e[SII]}$, so the abundances are well determined. Type~I PNe in the MW and LMC occupy the upper right of the diagram. Note that the plotted sample of SMC PNe lacks any Type I PNe. In this and following plots we have overlaid predicted PN abundances taken from \cite{K10} and \cite{M01} for the metallicities of the MW, LMC, and SMC, for initial masses from 1-6M$_{\odot}$ and 0.8-5M$_{\odot}$, respectively; solar values are shown by vertical and horizontal lines. The important point to note is that the models span the actual observations.

\begin{figure}[h]
\begin{center}
 \includegraphics[trim=1cm 0.8cm 0.8cm 0.5cm, angle=90, clip=true, totalheight=0.25\textheight]{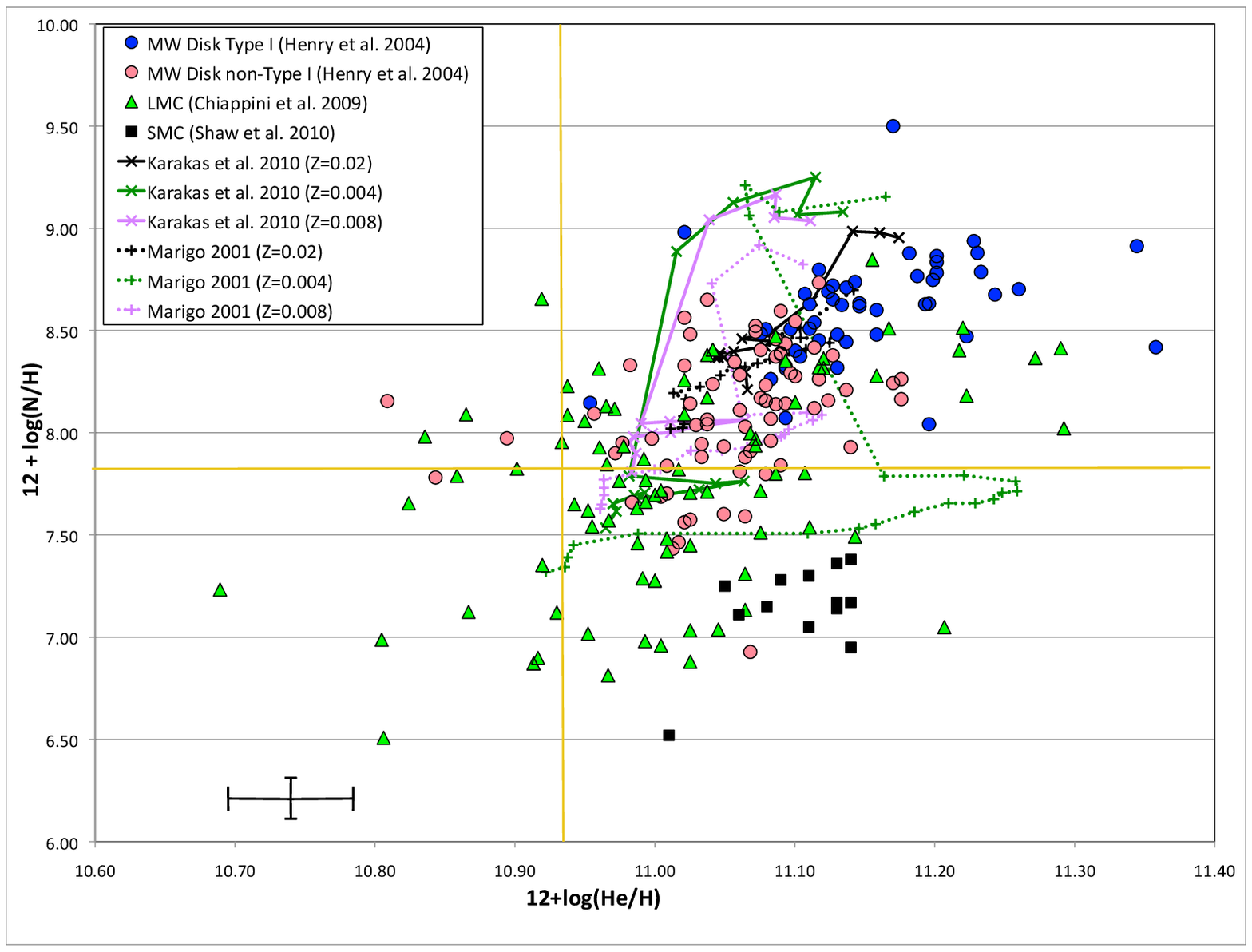} 
 \caption{N vs. He with model predictions. Type I PNe occupy the top right of the diagram.}
   \label{fig1}
\end{center}
\end{figure}

Fig.~2 plots C/O vs. O for various objects. The fairly narrow trend outlined by H~II regions, damped Lyman-$\alpha$ systems, halo stars, and disk stars is completely ignored by PNe, their wide swath testifying to their non-uniform C production.

\begin{figure}[h]
\begin{center}
 \includegraphics[trim=0cm 0cm 0cm 0cm, clip=true, totalheight=0.25\textheight]{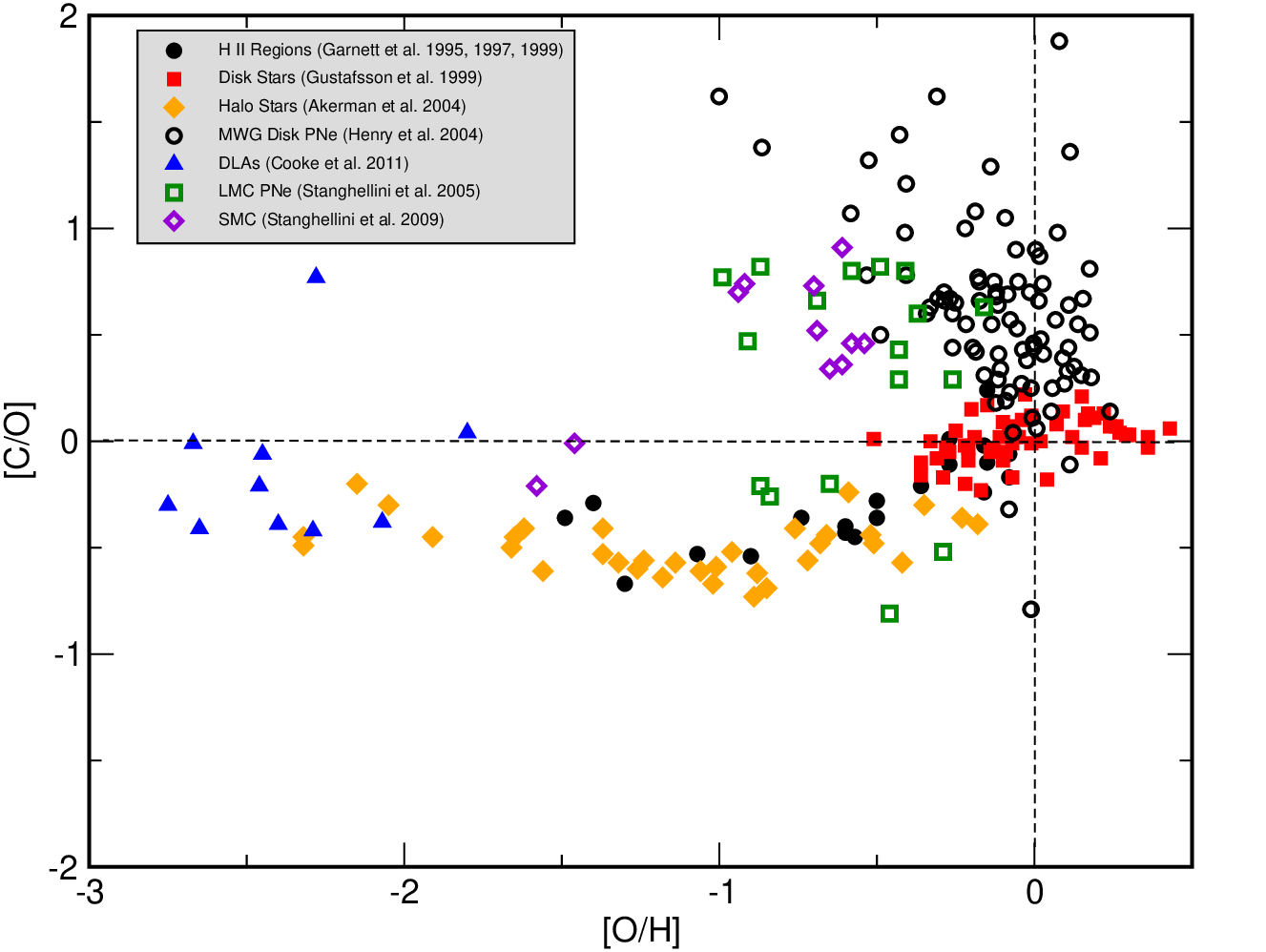} 

 \caption{C vs. O. Solar values are indicated by dashed lines.}
   \label{}
\end{center}
\end{figure}

In Fig.~3 we see N/O vs. O/H for the same samples as in Fig.~1. Note the trend of decreasing N/O with increasing O/H exhibited by the LMC PNe, and also by the MW Type I PNe; this is a sign of hot-bottom burning at the base of the convective envelope in massive TP-AGB stars, converting C and O into N. Note also that neither the SMC nor the MW non-Type I PNe exhibit this tendency.

\begin{figure}[h]
\begin{center}
 \includegraphics[trim=0.6cm 3cm 1cm 3cm, clip=true, totalheight=0.25\textheight]{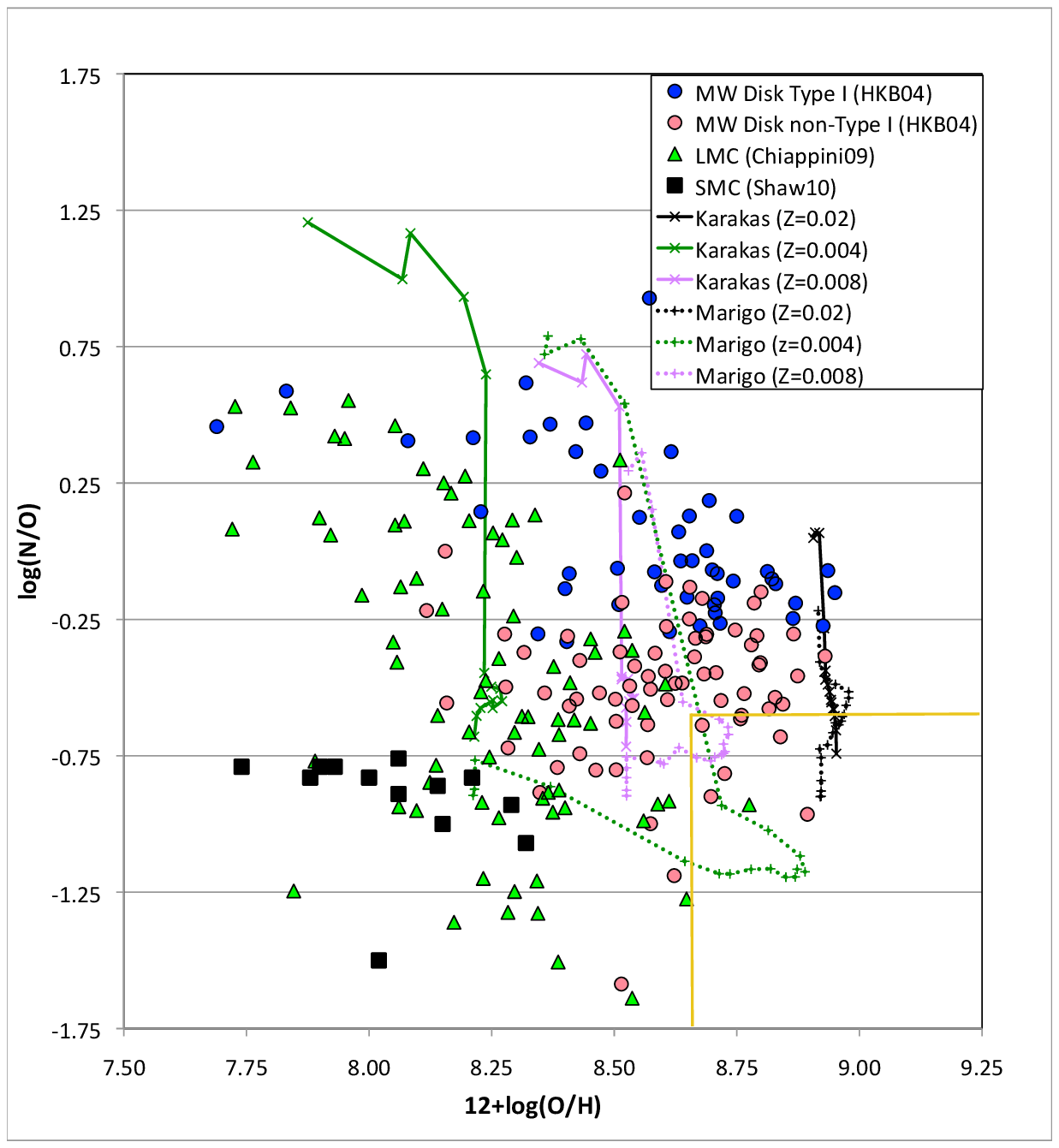} 

 \caption{N/O vs. O/H with solar values indicated by solid lines. Note the tendency  in MW Type~I and LMC PNe of decreasing O as N/O rises, a result of hot-bottom burning.}
   \label{}
\end{center}
\end{figure}

\subsection{Neon}

In Fig.~4 the Ne and O abundances are plotted for H~II regions and blue compact galaxies (H2BCG; Milingo et al. 2010) in addition to PNe in the MW disk and in M31's disk (\cite{K11}). The tight relation evident for the first two is followed, by PNe, but with more scatter. Model predictions from \cite{K10} are shown as well. The tendency for MW PNe to lie slightly, but systematically, above the H2BCG line could be interpreted as an indication that PNe are more enriched in Ne than H~II regions of the same metallicity (but see \S3.1). Also, Fig.~5  shows Ne/O vs. O for the same PNe samples as in Fig.~1 and there is no compelling evidence for a trend, which indicates either that O is not altered in LIMS nucleosynthesis at MW or MC metallicities, or that both Ne and O are changed similarly; \cite{MCI10} find the same result. \cite{WL07} concluded that Ne and O production only occurs for 12+log(O/H)$<$8, below even the SMC, indicating that the first possibility is likely the correct one. 

\begin{figure}[h]
\begin{center}
 \includegraphics[trim=0cm 1cm 0cm 0cm, clip=true, totalheight=0.25\textheight]{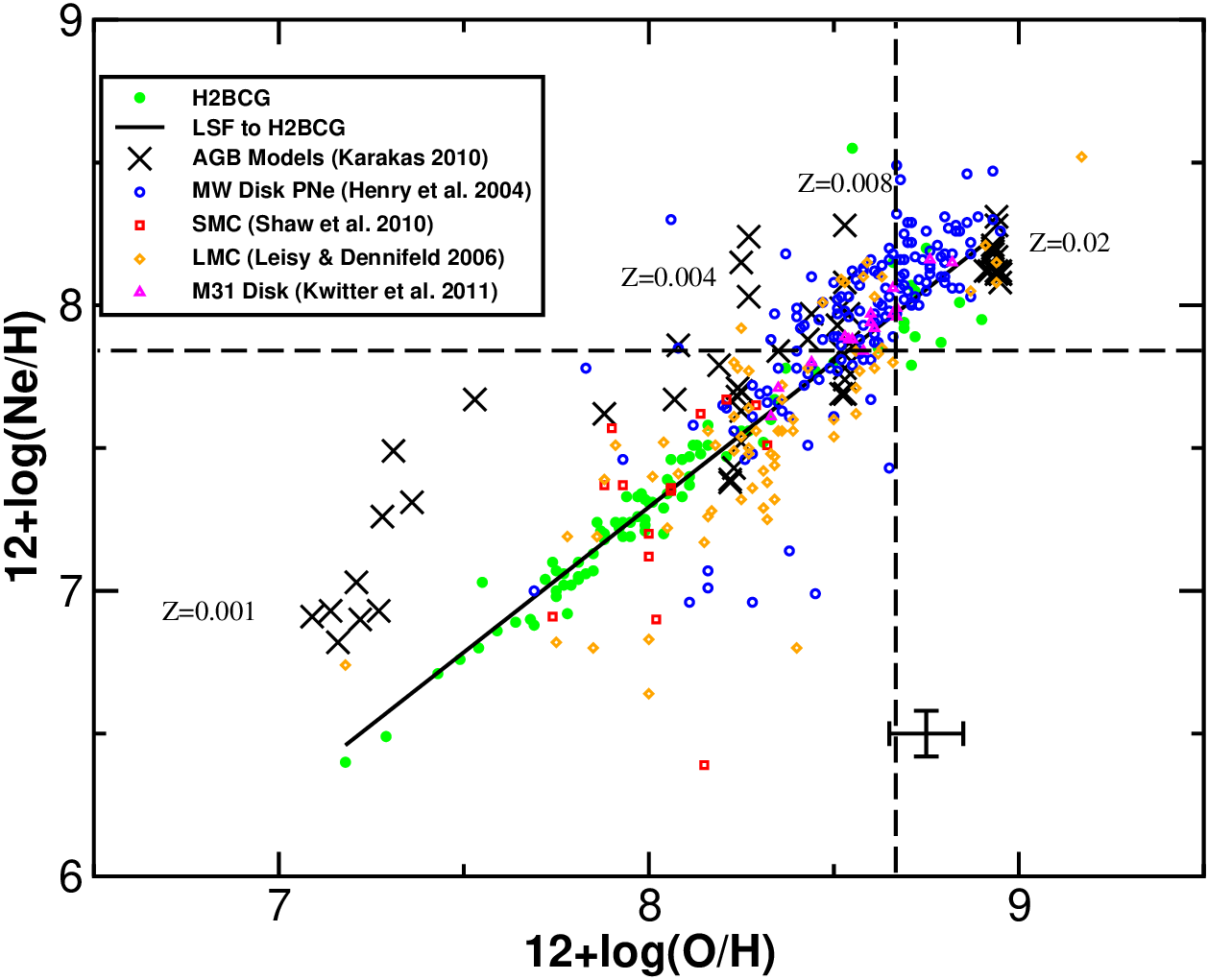} 

 \caption{Ne vs. O, with models. Solar values are indicated by dashed lines. For H2BCG references, see \cite{M10}.}
   \label{}
\end{center}
\end{figure}

\begin{figure}[h]
\begin{center}
 \includegraphics[trim=.8cm 2cm 1cm 2cm, angle=90, clip=true, totalheight=0.25\textheight]{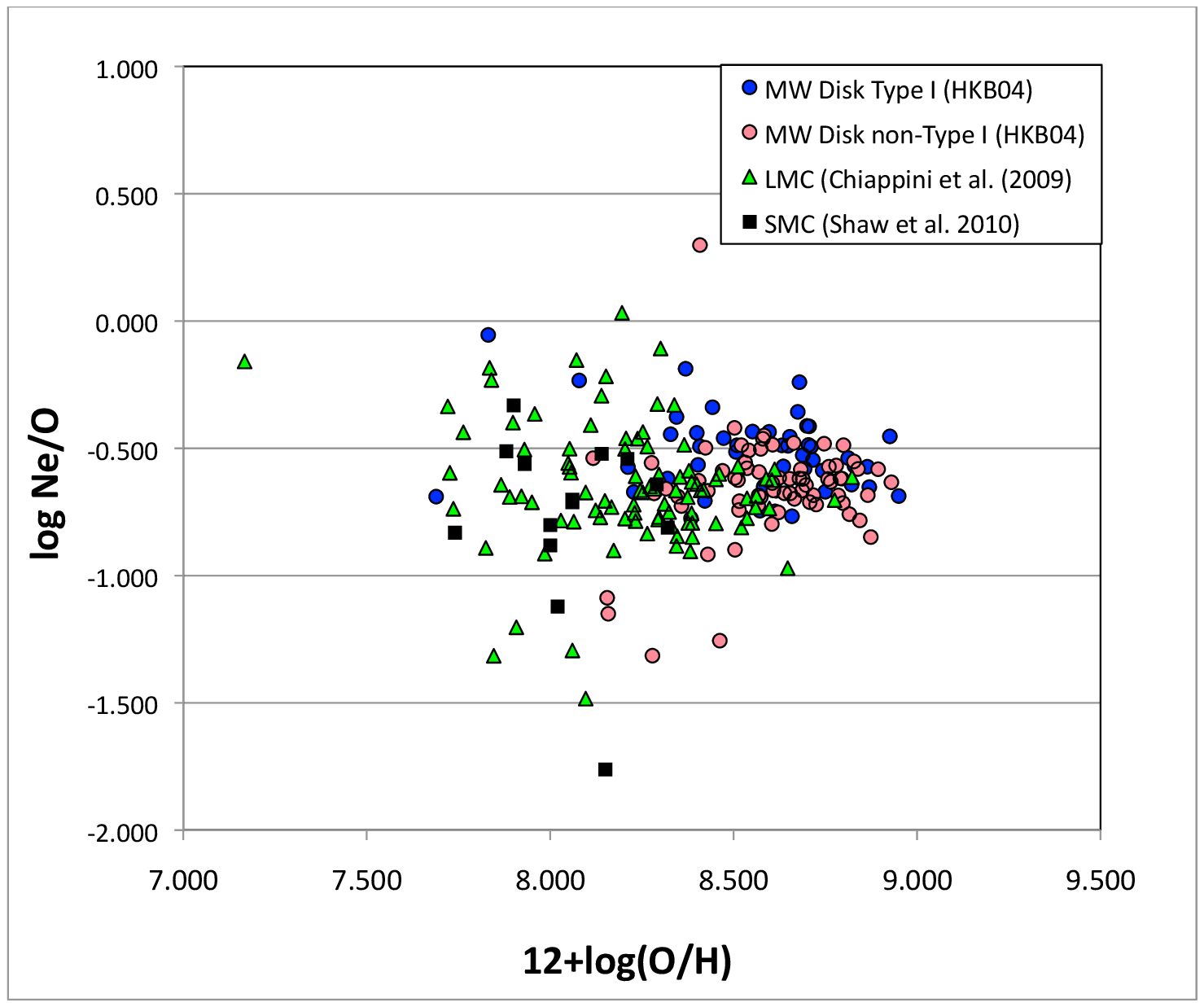} 

 \caption{Ne/O vs. O/H. The lack of a discernible trend implies that either O is not affected by LIMS evolution, or that both Ne and O are similarly altered.}
   \label{}
\end{center}
\end{figure}

\subsection{Sulfur and Argon}
In Fig.~6 we plot S vs. O, with other objects and model predictions as in Fig.~4. Also shown are the predictions (marked by X) of photoionization models for five MW PNe, confirming that they exhibit the sulfur anomaly. There is  a new possibility as to how the sulfur anomaly might be resolved, having to do with the existence of a S$^{+4}$ zone predicted by RHD models (see Jacob, this volume); we will be eager to see how it succeeds.

Fig.~7 shows Ar vs. O plotted for the same objects as in Fig.~4 for Ne. The trend set by the H2BCG sample is well followed by PNe, though the scatter is larger. The general concordance of Ne and Ar with O bolsters the view that these $\alpha$ elements do grow in lockstep, as predicted by nucleosynthesis arguments. 

\begin{figure}[h]
\begin{center}
 \includegraphics[trim=0cm 0cm 0cm 0cm, clip=true, totalheight=0.28\textheight]{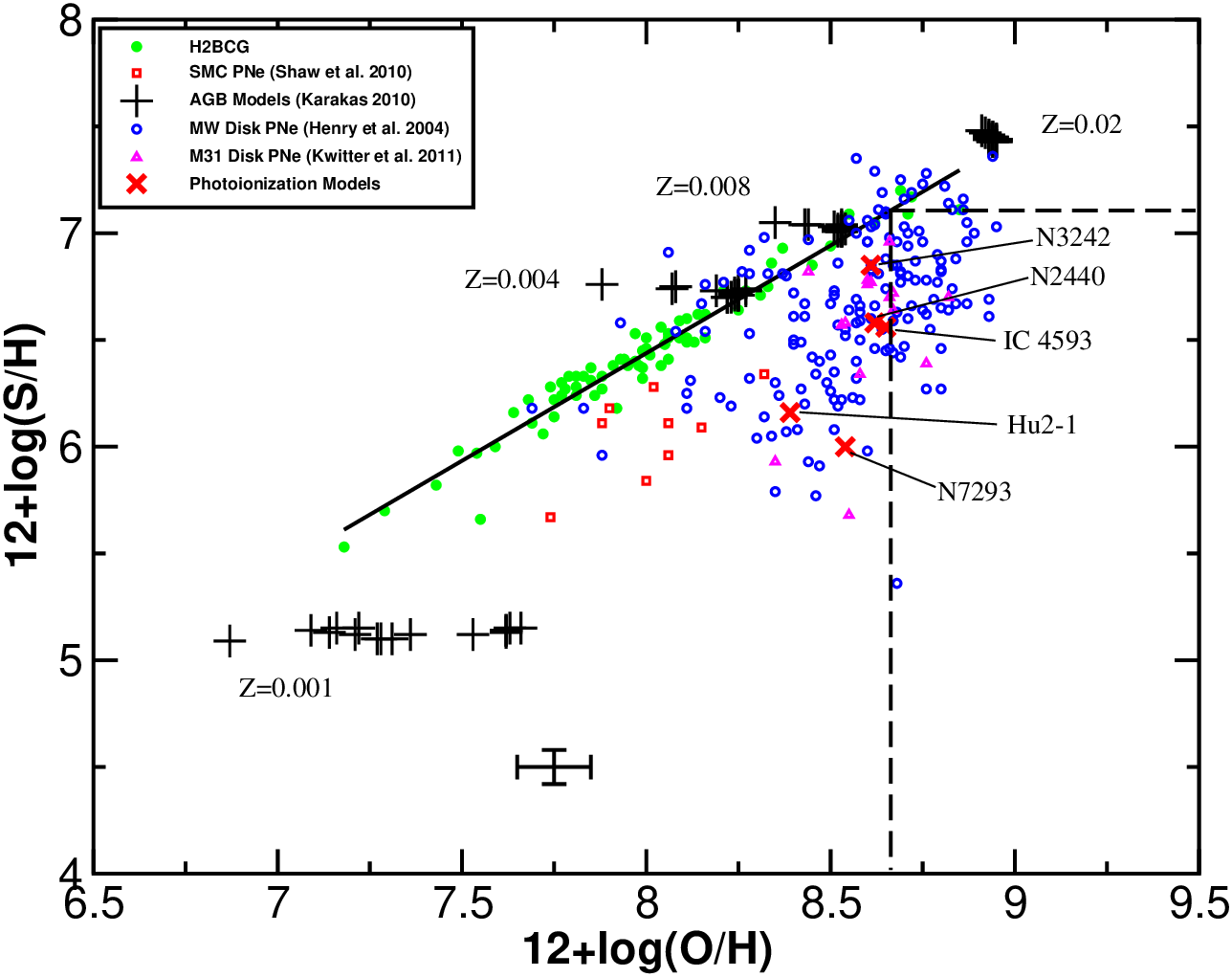} 

 \caption{S vs. O, with models. Solar values are indicated by dashed lines.}
   \label{}
\end{center}
\end{figure}

\begin{figure}[h]
\begin{center}
 \includegraphics[trim=0cm 0cm 0cm 0cm, clip=true, totalheight=0.25\textheight]{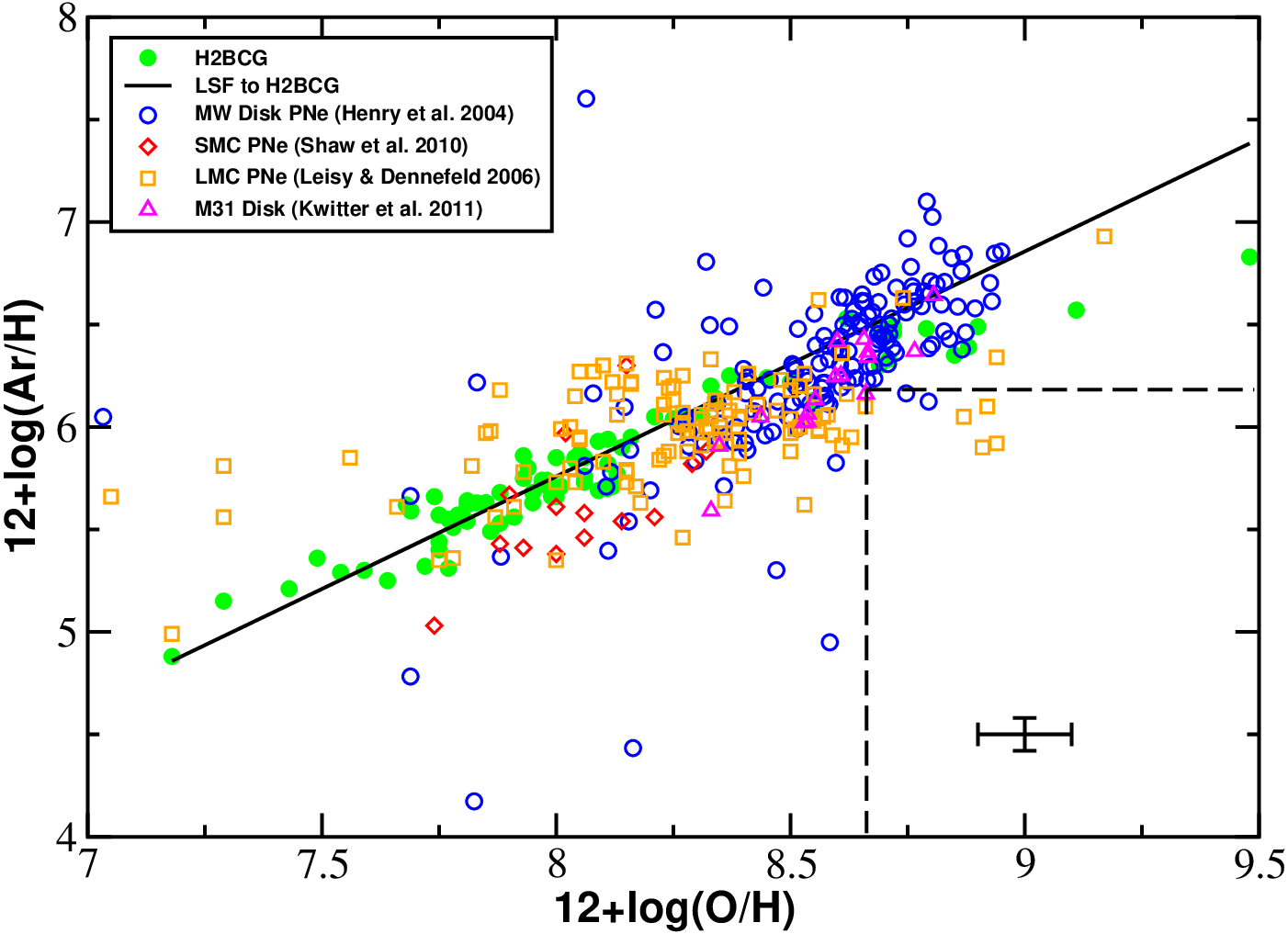} 

 \caption{Ar vs. O. Solar values are indicated by dashed lines.}
   \label{}
\end{center}
\end{figure}

\subsection{Morphological Correlations}
Striking correlations emerge when comparing different populations of PNe. \cite{MC10} show in their Fig.~1 that N and He abundances are significantly higher 
in asymmetric PNe compared with symmetric PNe; the opposite is true for C, implying that asymmetric PNe originate from systematically more massive stars. In addition, O, Ne, Ar, and S are all slightly higher in asymmetric PNe. This behavior is to be expected if the higher-mass progenitors of asymmetric PNe are younger and were born from a more-enriched ISM. For the MCs \cite{S09} found that in their samples, the C abundances in SMC PNe are significantly higher than in LMC PNe (their Fig.~13).

\section{Conclusions}
We end with a look ahead. Atomic data calculations for low temperatures and new elements and ions are now being carried out. A small group of investigators has begun a systematic look at the empirical abundance process as described in \S 3.1, with an eye toward evaluating and possibly even standardizing some of the procedure. Studies on the dependence of ICFs on metallicity, geometry and nebular ionization are also proceeding. Ongoing 2-D nebular abundance mapping and kinematic studies will enable 3-D nebular modeling, with the potential to solve, or at least to shed light on the abundance discrepancy. We anticipate exciting and illuminating new results at the next symposium. \newline

{\it Acknowledgments}: We are grateful to the NSF under grants AST-0806490 to Williams College and AST-0806577 to the University of Oklahoma, and to our respective institutions for support. We also thank the organizers for inviting us to present this review.


\begin{thebibliography}{}

\bibitem[Akerman et al. 2004]{A04}Akerman, C.J., Carigi, L., Nissen, P.E., Pettini, M., \& Asplund, M. 2004, \textit{A\&A}, 414, 931

\bibitem[Asplund et al.(2009)]{asplund09}Asplund, M., Grevesse, N., Sauval, A.-J., \& Scott, P. 2009, \textit{ARAA}, 47, 481

\bibitem[Bernard-Salas et al. (2008)]{BS08}Bernard-Salas, J., Pottasch, S., Morris, P., \& Houck, J. 2008, \textit{ApJ}, 672, 286

\bibitem[Cavichia et al. (2010)]{Ca10}Cavichia, O., Costa, R.D.D., \& Maciel, W.J. 2010, \textit{Rev. Mexicana AyA}, 47, 49

\bibitem[Chiappini et al. (2009)]{Ch09}Chiappini, C., G\'{o}rny, S., Stasi\'{n}ska, G., \& Barbuy, B. 2009, \textit{A\&A}, 494, 591

\bibitem[Clayton et al. (1989)]{CCM}Clayton, J.A., Cardelli, G.C., \& Mathis, J.S. 1989, \textit{ApJ}, 345, 245

\bibitem[Cooke et al. 2011]{Cooke11}Cooke, R., Pettini, M., Steidel, C.C., Rudie, G.C., \& Nissen, P. E. 2011, \textit{MNRAS}, in press

\bibitem[Delgado Inglada et al. (2009)]{D09}Delgado Inglada, G., Rodriguez, M., Mampaso, A., \& Viironen, K. 2009 {ApJ}, 694, 1335

\bibitem[Esteban et al. (2004)]{esteban04}Esteban, C., et al. 2004, \textit{MNRAS}, 355, 229

\bibitem[Fitzpatrick (1999)]{F99}Fitzpatrick, E., 1999, \textit{PASP}, 111, 63

\bibitem[]{}Garnett, D. R. et al. 1999, \textit{ApJ}, 513, 168

\bibitem[]{}Garnett, D. R. et al. 1995, \textit{ApJ}, 443, 64

\bibitem[]{}Garnett, D. R. et al. 1997, \textit{ApJ}, 481, 174

\bibitem[Gustafsson et al. 1999]{Gustafsson99}Gustafsson, B., Karlsson, T., Olsson, E., Edvardsson, B., \& Ryde, N. 1999, \textit{A\&A}, 342, 426

\bibitem[Henry et al. (2004)]{HKB04}Henry, R.B.C., Kwitter, K.B., \& Balick, B. 2004 \textit{AJ}, 127, 2284

\bibitem[Henry et al. (2010)]{H10}Henry, R.B.C., Kwitter, K.B., et al. 2010, \textit{ApJ}, 724, 748

\bibitem[Idiart et al. (2007)]{IMC07}Idiart, T.P., Maciel W.J., \& Costa, R.D.D. 2007, \textit{A\&A}, 472, 101

\bibitem[Johnson et al. 2006]{J06}Johnson, et al. 2006, in \textit{PNe in our Galaxy and Beyond}, Proc. IAU Symp. 234, p.\ 439

\bibitem[Karakas (2010)]{K10}Karakas, A.I. 2010, \textit{MNRAS}, 403, 1413

\bibitem[Kingsburgh \& Barlow (1994)]{KB94}Kingsburgh, R., \& Barlow, M. 1994 \textit{MNRAS}, 271, 257

\bibitem[Kwitter \& Henry 2001]{KH01}Kwitter, K.B., \& Henry, R.B.C. 2001, \textit{ApJ}, 562, 804

\bibitem[Kwitter et al. 2011]{K11}Kwitter, K.B., Lehman, E.M.M., Henry, R.B.C., \& Balick, B. 2011, {\it in preparation}

\bibitem[Leisy \& Dennefeld (2006)]{LD06}Leisy, P., \& Dennefeld, M. 2006, \textit{A\&A}, 456, 451

\bibitem[Liu et al. 2006]{L06}Liu, X.-W., Barlow, M.J., Zhang, Y., Baston., R.J., \& Storey, P.J. 2006 \textit{MNRAS}, 368, 1959

\bibitem[Maciel \& Costa (2010)]{MC10}Maciel, W.J., \& Costa, R.D. 2010, in Asymmetric PNe V, (San Francisco: ASP) 

\bibitem[Maciel et al. (2010)]{MCI10}Maciel, W.J.,et al. 2010, in Proc. 11th Symp. on Nuclei in the Cosmos, published online

\bibitem[Marigo (2001)]{M01}Marigo, P. 2001, \textit{A\&A}, 370, 194

\bibitem[Mendoza (1983)]{M83}Mendoza, C. 1983, in \textit{PNe}, Proc. IAU Symp. 103, (Dordrecht: Reidel), p.\ 143

\bibitem[Milingo et al. (2010)]{M10}Milingo, J.B., Kwitter, K.B., Henry, R.B.C., \& Souza, S.P. 2010, \textit{ApJ}, 711, 619

\bibitem[Pequignot \& Baluteau (1994)]{PB94}Pequignot, D., \& Baluteau, J.-P. 1994, \textit{A\&A}, 283, 593

\bibitem[Peimbert et al. 1992]{P92}Peimbert, M., Torres-Peimbert, \& Ruiz, M.-T. 1992, \textit{Rev. Mexicana AyA}, 24, 155

\bibitem[Porter et al. 2007]{P07}Porter, R., Ferland, G.J., \& McAdam, K.B. 2007, \textit{ApJ}, 657, 327


\bibitem[Pottasch \& Bernard-Salas (2010)]{PBS10}Pottasch, S.R., \& Bernard-Salas, J. 2010, \textit{A\&A}, 517, A95

\bibitem[Savage \& Mathis (1979)]{SM}Savage, B.D., \& Mathis, J.S. 1979, \textit{ARAA}, 17, 73

\bibitem[Sharpee et al. (2007)]{Sharpee}Sharpee, B.et al. 2007, \textit{ApJ}, 659, 1265 

\bibitem[Shaw \& Dufour 1995]{SD95}Shaw, R.A., \& Dufour, R.J. 1995, \textit{PASP}, 107, 896

\bibitem[Shaw et al. (2010)]{Shaw10}Shaw, R.A. et al. 2010, \textit{ApJ}, 717, 562

\bibitem[Stanghellini \& Haywood (2010)]{SH10}Stanghellini, A., \& Haywood, M. 2010, \textit{ApJ}, 714, 1096

\bibitem[Stanghellini et al. (2009)]{S09}Stanghellini, L., Lee, T.-H., Shaw, R.A., Balick, B., \& Villaver, E. 2009, \textit{ApJ}, 702, 733

\bibitem[Stanghellini et al. (2005)]{S05}Stanghellini, L., Shaw, R.A., \& Gilmore, D. 2005 \textit{ApJ}, 622, 294

\bibitem[Sterling \& Dinerstein (2008)]{SD08}Sterling, N.C., \& Dinerstein, H.L. 2008, \textit{ApJS}, 174, 158

\bibitem[Tsamis et al. 2003]{T03}Tsamis, Y.G., Barlow, M.J., Liu, X.-W., Danziger, I.J., \& Storey, P.J. 2003, \textit{MNRAS}, 245, 186

\bibitem[Tsamis et al. 2008]{T08}Tsamis, Y. G., Walsh, J.R., \& Pequignot, D. 2008, \textit{MNRAS}, 386, 22

\bibitem[Torres-Peimbert \& Peimbert 2003]{TPP03}Torres-Peimbert, S., \& Peimbert, M. 2003, in Proc. IAU Symp. 209, p.\ 363

\bibitem[Wang \& Liu (2007)]{WL07}Wang, W., \& Liu, X.-W. 2007, \textit{MNRAS}, 381, 669 

\end{thebibliography}
\end{document}